\documentclass[journal]{IEEEtran}
\usepackage[T1]{fontenc}
\usepackage{graphicx}
\usepackage{textcomp}
\usepackage{xcolor}
\usepackage{amssymb,amsmath,amsfonts,amsthm}
\usepackage{mathrsfs}
\usepackage{cite}
\usepackage{array}
\usepackage{tabularx}
\usepackage{color}
\usepackage{subfigure}
\usepackage{mathtools}
\usepackage{tikz,pgf}
\usetikzlibrary{calc,positioning,mindmap,trees,decorations.pathreplacing}
\usepackage{graphicx}
\usepackage{bbm}
\usepackage{bm}
\usepackage[utf8]{inputenc}
\usepackage{algorithm}
\usepackage{algcompatible}

\newcommand{\matr}[1]{\mathbf{#1}}
\newcommand{\vect}[1]{\bold{#1}}

\usepackage[colorinlistoftodos,prependcaption,backgroundcolor=black!5!white,bordercolor=red]{todonotes}

\begin{document}
	
	\title{Distributed Consensus in Wireless Networks with Probabilistic Broadcast Scheduling}
	
	\author{{Daniel Pérez Herrera, Zheng Chen and Erik G. Larsson}
		
		\thanks{The authors are with the Department of Electrical Engineering, Link\"{o}ping University, Link\"{o}ping, SE-58183 Sweden. E-mail: \{daniel.perez.herrera, zheng.chen, erik.g.larsson\}@liu.se. This work was supported by Zenith, ELLIIT, and the KAW foundation.}
	}
	
	\maketitle
	
	\begin{abstract}
		We consider distributed average consensus in a wireless network with partial communication to reduce the number of transmissions in every iteration/round. Considering the broadcast nature of wireless channels, we propose a probabilistic approach that schedules a subset of nodes for broadcasting information to their neighbors in every round.
		We compare several heuristic methods for assigning the node broadcast probabilities under a fixed number of transmissions per round. Furthermore, we introduce a pre-compensation method to correct the bias between the consensus value and the average of the initial values, and suggest possible extensions for our design. Our results are particularly relevant for developing communication-efficient consensus protocols in a wireless environment with limited frequency/time resources.    
	\end{abstract}
	
	
	\section{Introduction}
	Distributed consensus deals with the problem of reaching an agreement among networked agents on a certain quantity of interest that depends on the initial states of all agents. The interaction rule that specifies the information exchange between an agent (node) and its neighbors is known as consensus protocol or algorithm \cite{olfati2007consensus}.
	Traditional studies of consensus problems usually assume a fixed topology with perfect communication between the nodes in the network. For this ideal case, the convergence conditions and convergence speed of discrete-time average consensus are provided in \cite{xiao2004fast} and \cite{olshevsky2009convergence}. As an extension to the main theme, different methods can be used to accelerate convergence, such as extrapolation \cite{kokiopoulou2007accelerating} and stepsize adjustment \cite{mosquera2010stepsize}.
	For varying topologies, \emph{necessary and sufficient} conditions to reach consensus over random networks are presented in \cite{tahbaz2008necessary}, using ergodicity and probabilistic arguments. 
		
	Conventional works on consensus algorithms tend to rely on idealized assumptions (e.g., instantaneous, error-free) on the communication links between connected nodes.
	Consensus algorithms may suffer from imperfect communication between the nodes due to, for example, link failures \cite{patterson2010convergence}, \cite{kar2008distributed}, \cite{fagnani2008randomized}. To deal with the missing information, two compensation (biased and balanced) methods are presented in \cite{fagnani2009average}. The main difference between these methods is how the weights of the failed links are re-distributed among the successful links.  
	Both algorithms converge almost surely and in the mean-square sense to a value that generally is not equal to the average of the initial values.

	In addition to the conventional consensus protocols where every node communicates with all their neighbors in every iteration,  gossip algorithms can also be used to achieve consensus with randomized and asynchronous communication \cite{dimakis2010gossip}, \cite{boyd2005gossip}. 
	A key feature of these algorithms is the asynchronous time model, where information is exchanged between a subset of nodes. This information is processed by the receiving nodes to compute local update via a linear combination of its own value and the received one. There are different variants of gossip algorithms, such as geographic gossip \cite{dimakis2008geographic}, randomized gossip \cite{boyd2006randomized} and broadcast gossip \cite{aysal2009broadcast}. Particularly, broadcast gossip is more relevant to the implementation of gossip algorithms in wireless networks, where the transmission of one node can reach any other nodes within its communication range. 
		
	Unlike the case with broadcast gossip, where the nodes update their states immediately after receiving information from another node, we propose a probabilistic broadcast-based method to schedule a subset of nodes for broadcasting in every communication round. The nodes update their values after receiving information from all the scheduled broadcasting nodes. Since only a subset of nodes is selected for transmission in every round, the number of transmission slots needed in every round is reduced as compared to the full communication case. To compensate for the lack of information with our partial communication approach, we use a biased compensation method similar to the one proposed in \cite{fagnani2009average}. Our objective is to achieve fast convergence with fewer transmission slots while maintaining small deviation from the average of the initial values. We investigate several heuristic methods for assigning the broadcast probabilities of the nodes, combined with a pre-compensation mechanism to eliminate the bias caused by unbalanced information exchange.
	
	\section{System Model}
	
	\subsection{Graph Model}
	We consider a wireless network modeled by an undirected and connected graph $\mathcal{G}=(\mathcal{N},\mathcal{E})$, where $\mathcal{N}$ is the set of nodes and $\mathcal{E}$ is the set of edges. The total number of nodes is $N$. Each edge $\{i,j\}\in \mathcal{E}$ indicates the existence of a link between the pair of nodes $i$ and $j$. The set of neighbors of node $i$ is $\mathcal{N}_i = \{ j|\{i,j\}\in \mathcal{E}\}$.\footnote{In a wireless network, the notion of connectivity is not strictly defined since every node can hear the transmission of other nodes in the network. We consider a simplified model where a link between two nodes does not exist when the distance is larger than some predefined threshold.  }
	
	The connectivity of the graph can be represented by its adjacency matrix $\matr{A}$, with elements $a_{ij}=1$ if $\{i,j\}\in \mathcal{E}$, and $a_{ij}=0$ otherwise. Another matrix representation of the graph is the Laplacian matrix, defined as $\matr{L} = \matr{D} - \matr{A}$, where $\matr{D}$ is a diagonal matrix whose $i$-th entry is $d_{i}=|\mathcal{N}_i|$.
	
	\subsection{Consensus Algorithm}
	\label{sec:consensus}	
	We consider the problem of distributed average consensus, where the goal is for all nodes in the network to reach consensus on the arithmetic mean of their initial values, by only communicating with their neighbors. 
	To accomplish this, we use the following distributed linear iteration
	\begin{equation}
		\vect{x}(t+1) = \matr{W}\vect{x}(t),
		\label{eq1}
	\end{equation} 
	where $\vect{x}(t) = [x_1(t),x_2(t),...,x_N(t)]^T$ contains the values of every node in iteration $t\geq 0$. $\matr{W}$ is the mixing matrix, defined as $\matr{W}=\matr{I}-\epsilon \matr{L}$, where $\matr{I}$ is the identity matrix and $\epsilon \in \mathbb{R}$ is the step size. Let $\Delta(\mathcal{G})$ be the maximum degree of the graph $\mathcal{G}$. When $0 < \epsilon < 1/\Delta(\mathcal{G})$, the following are necessary and sufficient conditions for $\vect{x}(t)$ to converge to the average of the initial values \cite{xiao2004fast}:	
	\begin{enumerate}
		\item $\matr{W}\vect{u} = \vect{u}$,
		\item $\vect{u}^T\matr{W}=\vect{u}^T$,
		\item $\rho(\matr{W}-\vect{u}\vect{u}^T/N) < 1$,
	\end{enumerate}
	where $\rho(\cdot)$ denotes the spectral radius of a matrix and $\vect{u}$ the all 1's column vector. Note that since $\matr{W}$ is symmetric, conditions $1$ and $2$ are equivalent. Condition 3 is also related to the convergence speed, since minimizing the second largest eigenvalue of $\matr{W}$ can speed up convergence of distributed averaging algorithms \cite{xiao2004fast}. 
	
	\section{Partial Communication with Probabilistic Broadcast Scheduling}	
	
	Distributed consensus algorithms rely on an iterative process where in each iteration, every node must communicate with all its neighbors before the updating step in \eqref{eq1}. 
	The communication design of consensus algorithms in wireless networks is non-trivial due to the ``many-to-many'' communication topology. If all nodes transmit information simultaneously using the same frequency-time resources, then all signals will be garbled. One simple way to avoid interference between concurrent transmissions is to assign orthogonal resources to each transmitting node. For instance, each node occupies one dedicated time slot for broadcasting its information to the neighbors. In this way, it takes $N$ times slots in total for the entire network to complete information fusion in one consensus iteration with full communication.  
	
	We refer to one \textit{communication round} as one iteration interval during which all nodes exchange information with the neighbors and update their state values, and one \textit{transmission slot} as the time interval during which one node broadcasts its value. With full communication over the network, after $J$ iterations of the consensus algorithm, the total number of required transmission slots is $NJ$. 
	
	To reduce the communication costs (time slots), we consider a partial communication design where in every round we only select a subset of nodes for broadcasting their information. 
	Let $\vect{p} = [p_1,p_2,...,p_N]^T$ be a vector that contains the broadcast probabilities of the nodes with $\sum_{i=1}^{N} p_i = K$ and $K<N$. This means that in every communication round, $K$ nodes are selected on average for broadcasting. Therefore, the average number of transmission slots for completing $J$ iterations is $KJ<NJ$.
	
	Since in every iteration, some nodes are not scheduled for broadcasting their values, we compensate the missing information by using the biased compensation method proposed in \cite{fagnani2009average}. The idea is to use a new mixing matrix $\overline{\matr{W}}(t)$ in every round $t$, whose elements are given by
	\begin{equation}
		\overline{w}_{ij}(t) = \left\{ \begin{array}{lcc}
			w_{ij}v_j(t), &   \text{if}  & j\neq i.\\
			\\ 1-\sum_{k=1, k\neq i}^{N} w_{ik}v_k(t), &  \text{if} & j = i. \\
		\end{array}
		\right.
		\label{expected_W}
	\end{equation}
	Here, $v_j(t)=1$ if node $j$ is selected for broadcasting in the $t$-th communication round, and $v_j(t)=0$ otherwise. With our probabilistic scheduling approach, we have $p_i=\mathbb{E}[v_i(t)]$.
	Though the biased compensation method ensures convergence in mean square, in general the consensus value is not equal to the average of the initial values, since the new mixing matrix is only row stochastic but not necessarily doubly stochastic.  
	
	\subsection{Heuristic Designs for Broadcast Probability Vector}	
	Intuitively, the broadcast probabilities should be determined in a way that can reflect the ``importance'' of the nodes in a network. As a first approach, we consider different methods of creating the broadcast probability vector based on commonly used metrics for network connectivity, such as degree, PageRank and betweenness centrality \cite{latora2017complex}. 
	
	\subsubsection{Degree-based method}
	Degree centrality of a node is the number of neighbors of the node: the degree of node $j$ is $d_{j}=|\mathcal{N}_j|$.
	We can choose the broadcast probability of node $i$ as
	\begin{equation}
		p_i = \min\{1, \gamma d_{i}\}
	\end{equation}
	where the constant $\gamma$ is chosen such that $\sum_{j=1}^{N} p_j = K$, for a given value of $K$.
	
	\subsubsection{PageRank-based method}
	PageRank was first proposed to rank webpages using the hyperlink network structure of the web. It can be used in any network as a measure of the node importance. The PageRank-based method gives us a probability vector $\vect{p}$ in the same way, but using the PageRank centrality of the nodes instead of their degrees.
	
	\subsubsection{Betweenness-based method}
	Betweenness centrality of a node is defined as the fraction of shortest paths between pairs of other nodes that pass through it.  
	Let $b_i$ represent the betweenness centrality of node $i$. Since $b_i$ is zero if node $i$ has degree one, a small positive value is added to this metric, i.e., the new score $\tilde{b}_i$ of node $i$ is:
	\begin{equation}
		\tilde{b}_i = b_i + \beta, 
	\end{equation}
	where $\beta$ is a small positive number.
	The probability vector $\vect{p}$ is then created in the same way as for the degree-based method, but using this new score instead of the node degree.

	\subsection{Bias Correction}
	\label{sec:bias-correction}
	
	With our probabilistic scheduling of broadcasting nodes in every round, the matrix $\overline{\matr{W}}(t)$ is row stochastic with positive diagonals (as long as $p_i>0, \forall i$). As a consequence of Theorem 2.1 in \cite{fagnani2009average}, the product $\prod_{t=0}^{J} \overline{\matr{W}}(t)$ converges almost surely, when $J\to\infty$, to a random rank-one matrix, $\vect{u}\boldsymbol{\alpha}^T$, for some vector $\boldsymbol{\alpha}=[\alpha_1,\alpha_2,...,\alpha_N]^T$. In general, a bias will be introduced, meaning that the consensus value will not be the average of the initial values.
	There are several existing bias correction approaches in the literature of distributed consensus, such as the corrective consensus in \cite{asymcorrective}, and the re-scaling method in \cite{xi2018linear}. The method proposed in \cite{xi2018linear} is not directly applicable in our system because of the time-varying communication topology. However, it is possible to 
	correct the bias by using a pre-compensation method as follows.

	For a given topology $\mathcal{G}=(\mathcal{N},\mathcal{E})$, a set of initial values $\{x_i(0)\}_{i=1}^N$ and a scheduling probability vector $\vect{p}$:
	\begin{enumerate}
		\item Run $N$ consensus processes with probabilistic scheduling of broadcast transmissions. In the $i$-th consensus process, the initial value of every node is zero, except for node $i$, whose initial value is one. In this case all nodes will converge to $\alpha_i$, since $\prod_{t=0}^{\infty} \overline{\matr{W}}(t)\vect{e}_i =\vect{u}\boldsymbol{\alpha}^T\vect{e}_i=\alpha_i\vect{u}$, where $\vect{e}_i$ is the $i$-th column of the $N\times N$ identity matrix. 
		\item Pre-compensate the initial values: for node $i$, the new initial value becomes $\tilde{x}_i(0) = \frac{x_i(0)}{\alpha_i N},   i = 1,...,N$. 
		\item Run another consensus algorithm with initial node values $\{\tilde{x}_1(0),\tilde{x}_2(0),...,\tilde{x}_N(0) \}$.
		This consensus process will converge to  $\frac{1}{N}\sum_{i=1}^{N}x_i(0)$.
	\end{enumerate}
		
	Importantly, the same seed must be used for the generation of (pseudo-)random numbers in the consensus mechanisms in Steps 1 and 3. This can be achieved, for example, by using a source of shared randomness. 
	Note that the pre-compensation weights computation (Step 1)   before the actual consensus process (Step 3) requires extra communication and computation. Specifically, it will consume $J KN$ transmission slots, where $J$ is the number of iterations in Step 1. However, this extra cost is incurred only once, since the pre-compensation weights can be reused for any number of future consensus processes, as long as the network topology and the probability vector remain the same.

	\subsection{Possible Extensions}	
	\label{sec:optimization}
	\subsubsection{Numerical Optimization of Broadcast Probabilities}
	As mentioned in Sec.~\ref{sec:consensus}, for a fixed graph, we can optimize the second largest eigenvalue of the mixing matrix to achieve fast convergence as shown in \cite{xiao2004fast}. However, following our approach with partial communication, $\overline{\matr{W}}(t)$ is a time-varying random matrix whose elements are determined by the scheduling decision in every communication round. It is generally difficult to analyze the eigenvalues of $\prod_{t=0}^{\infty}\overline{\matr{W}}(t)$. Therefore, we focus on the expectation of $\overline{\matr{W}}(t)$ and optimize the probability vector $\vect{p}$  by minimizing the second largest eigenvalue of $\mathbb{E}[\overline{\matr{W}}(t)]$. The elements of $\mathbb{E}[\overline{\matr{W}}(t)]$ are:
	\begin{equation}
		\mathbb{E}[\overline{w}_{ij}(t)] = \left\{ \begin{array}{lcc}
			w_{ij}p_j, &  \text{if}  & j\neq i. \\
			\\ 1-\sum_{k=1, k\neq i}^{N} w_{ik}p_k, &  \text{if} & j = i. \\
		\end{array}
		\right.
		\label{expected_W}
	\end{equation}
	
	The optimization problem can be defined as: 
	\begin{equation}
		\begin{aligned}
			\min_{\vect{p}} \quad & \rho(\mathbb{E}[\overline{\matr{W}}(t)]-\vect{u}\vect{u}^T/N)\\
			\textrm{s.t.} \quad & \sum_{i=1}^{N}p_i = K\\
			& 0\leq p_i \leq 1, \quad i=1,...,N\\
		\end{aligned}
	\label{eq:prob-opt}
	\end{equation}
	for a given value of $K$.

	To the best of our knowledge, there is no tractable expression that can characterize the relation between the second largest eigenvalue of  $\mathbb{E}[\overline{\matr{W}}(t)]$ and the broadcast probability vector $\vect{p}$.
	It is possible to apply some derivative-free optimization method, such as Simultaneous Perturbation Stochastic Approximation (SPSA), for optimizing the broadcasting probabilities \cite{spall1998overview},\cite{sadegh1997constrained}. The convergence analysis of SPSA has been established in the literature with different assumptions on the objective function. However, due to the unknown structure of the objective function in \eqref{eq:prob-opt}, the convergence to the optimal solution in this case is not guaranteed.
	Another challenge of using this method for the eigenvalue optimization problem is that some broadcast probabilities can be zero depending on the network topology and the given $K$, which is undesirable. In practice, this can be handled by imposing a minimum value of the probabilities.

	\subsubsection{Alternative Bias Correction Methods}
	Another possible bias correction method is to apply the corrective consensus mechanism proposed in \cite{asymcorrective}. In that mechanism, a new set of auxiliary variables $\phi_{ij}(t)$ is introduced to represent the amount of change that node $i$ has made to its state value $x_i(t)$, due to the information received from its neighbor $j$ at iteration $t$. 
	Periodically, one corrective iteration takes place, where every node $i$ transmits $\phi_{ij}(t)$ to the corresponding neighbor $j$ and computes $\Delta_{ij}(t)=\phi_{ij}(t)+\phi_{ji}(t)$. This new set of variables $\Delta_{ij}$ represents the bias accumulated in both directions and it is used to correct the values of $x_i$ and $\phi_{ij}$.
	Note that this bias correction method introduces extra communication overhead. In every corrective iteration, $\sum_{i=1}^{N}d_i$ transmission slots are needed for the exchange of $\phi_{ij}$. Another issue with this method is that convergence is not guaranteed, even for a complete graph. (The convergence criteria given in \cite{asymcorrective} assume retransmissions).
 
	\section{Simulation Results}
	
	For simulations, we generate one instance of an Erd\"{o}s-Rényi random graph with $N=100$ nodes, that is undirected and connected. The initial values of the nodes are created by using a normal distribution with zero mean and unit variance. Every plot is obtained by averaging 10 realizations with the same graph but different initial values. The mixing matrix $\matr{W}=\matr{I}-\epsilon \matr{L}$ is computed for $\epsilon=\frac{1}{\Delta(\mathcal{G})+1}$. 
	
	\subsection{Broadcast Probability Vector Design}
	\label{sec:heuristic}
\begin{figure}[ht!]
		\centering
		\includegraphics[width=\columnwidth]{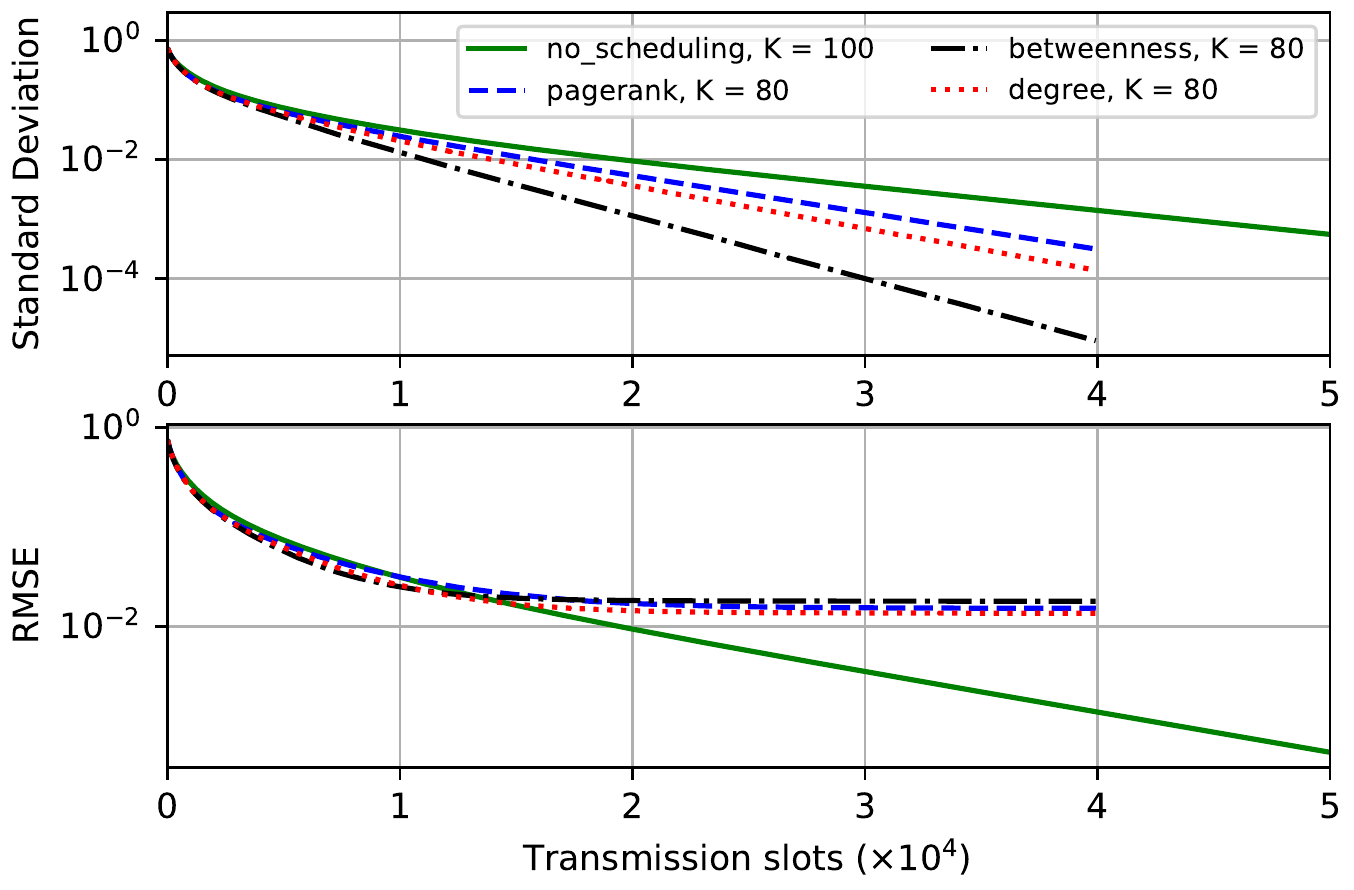}
		\caption{Comparison of the standard deviation of the nodes values and its distance from true average (RMSE) for different heuristic methods.}
		\label{fig:1A}
	\end{figure}

	First, we show the performance of our partial communication design with different heuristic choices for the broadcast probabilities. We choose $K=80$, which corresponds to $80\%$ of the nodes scheduled for broadcasting their values in every communication round. The results are presented in Fig.~\ref{fig:1A}. We observe a clear gain in terms of convergence speed (measured in transmission slots) with our partial communication design, especially with betweenness-based method. However, the converged value is not equal to the average of the initial values, which implies a tradeoff between the convergence speed and the bias in the consensus value. 
	
	\subsection{Partial Communication with Bias Correction}
\begin{figure}[ht!]
		\centering
		\includegraphics[width=\columnwidth]{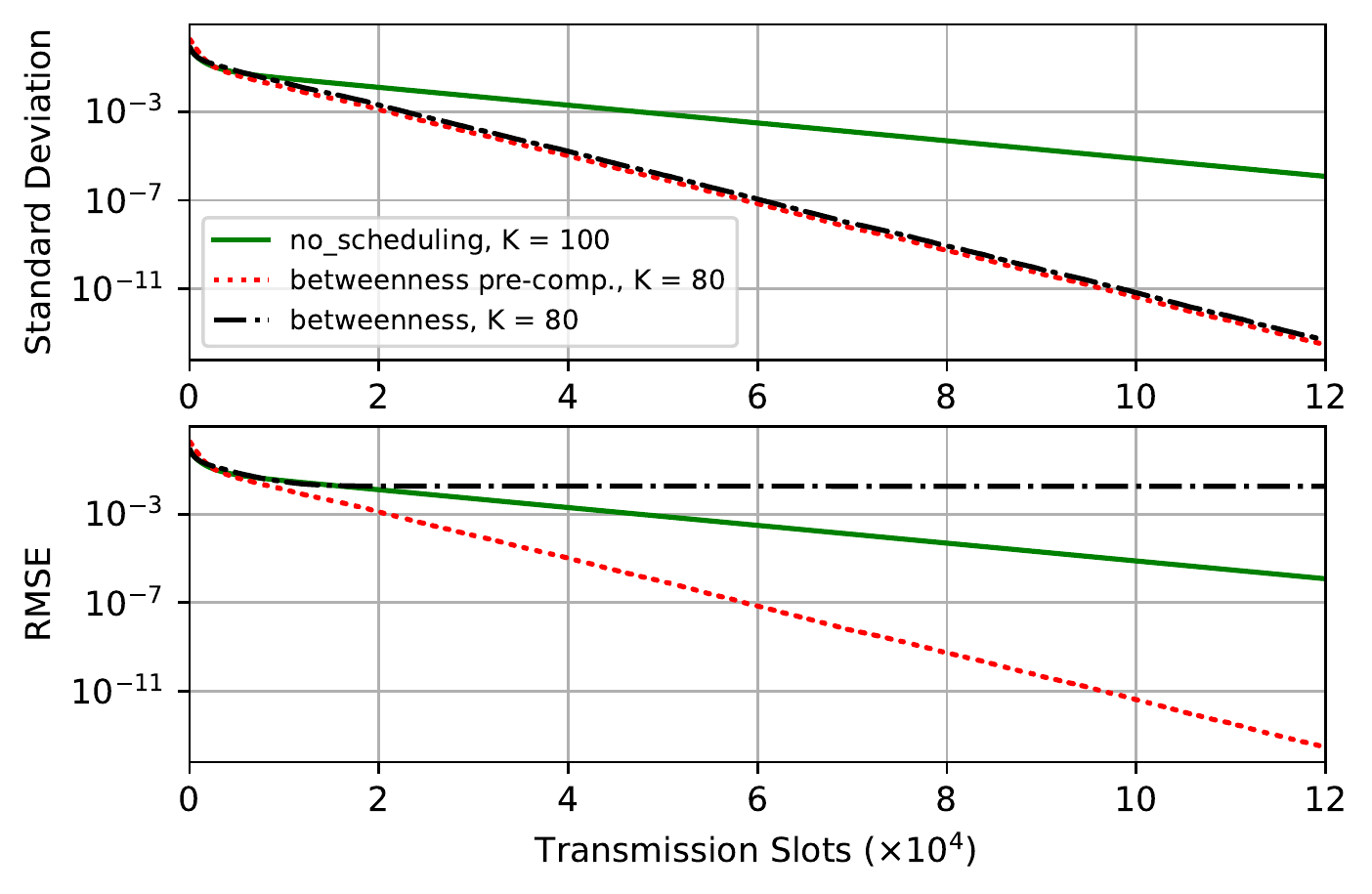}
		\caption{Performance with the proposed bias correction mechanism,  using the betweenness-based probability vector design.}
		\label{fig:corrective}
	\end{figure}

	To deal with the bias introduced by our partial communication design, we implement the pre-compensation mechanism introduced in Sec.~\ref{sec:bias-correction}. The results are presented in Fig.~\ref{fig:corrective}. We can see that by introducing a pre-compensation step before the actual consensus process, we can  eliminate the bias in the consensus value while maintaining the advantage of our proposed design in reducing the communication costs.

	\subsection{Potential Improvement by Using SPSA}
\begin{figure}[ht!]
		\centering
		\includegraphics[width=\columnwidth]{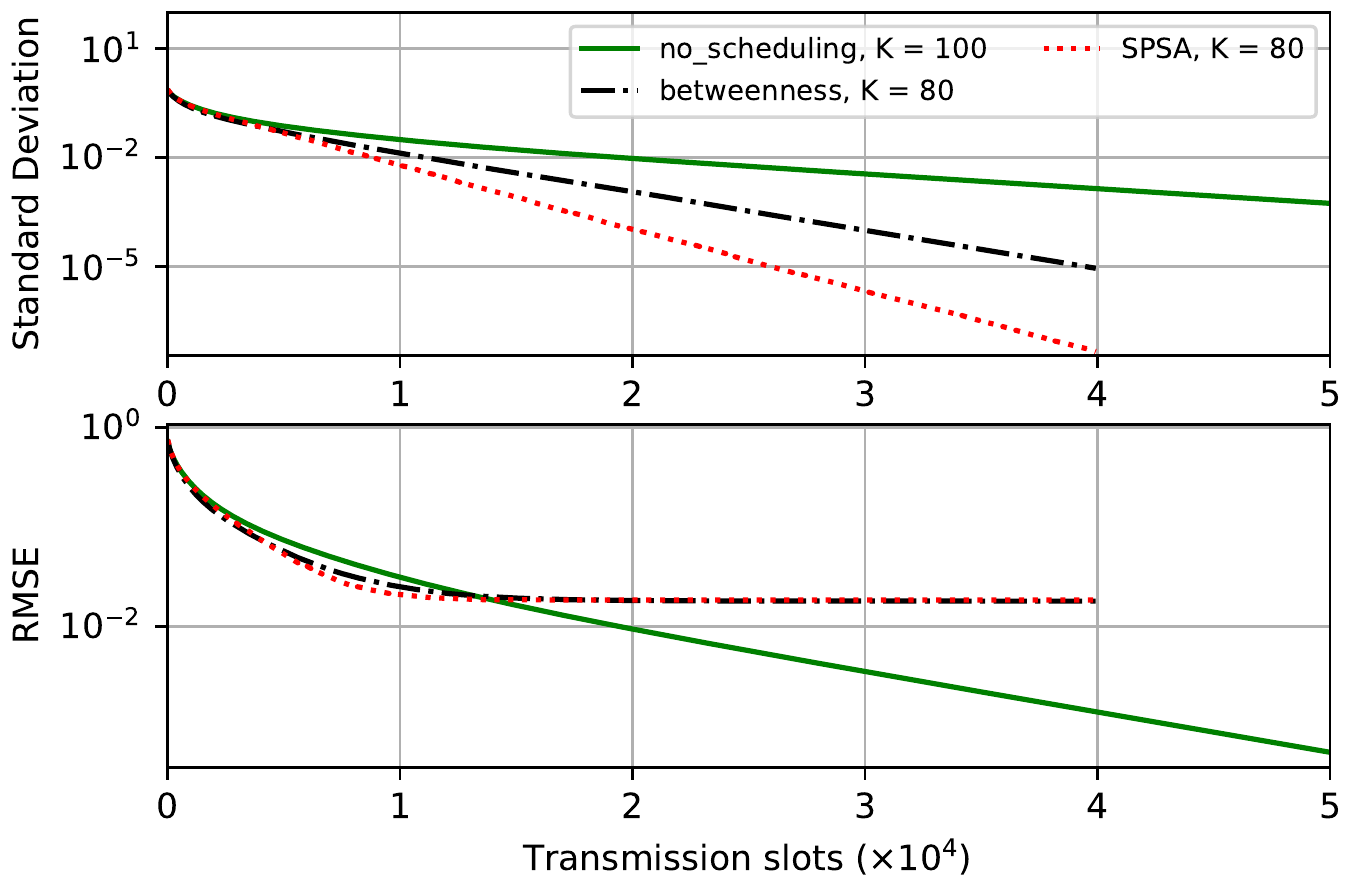}
		\caption{Comparison of the standard deviation of the nodes values and the RMSE for no scheduling, betweenness-based and SPSA-based methods.}
		\label{fig:1B}
	\end{figure}

	Finally, we show the performance of our partial communication design with the scheduling probabilities obtained with SPSA. 
	In Fig.~\ref{fig:1B}, we compare the SPSA-based method and the heuristic design of the probability vector using betweenness centrality, as detailed in Sec.~\ref{sec:heuristic}. 	
	
	{We observe that with the SPSA-based method, we can achieve faster convergence as compared to the heuristic design, while keeping similar RMSE. This result is well supported by the fact that reducing the second largest eigenvalue of the expected mixing matrix improves convergence speed.

	\section{Conclusion}
	We proposed a partial communication design for distributed average consensus over wireless networks using probabilistic broadcast scheduling. 
	A trade-off between convergence speed in terms of transmission slots and the bias in the consensus value was observed from simulations. Several heuristic methods for assigning the node broadcast probabilities were proposed, as well as a pre-compensation mechanism for bias correction. We concluded that distributed consensus algorithms in wireless networks can benefit from partial communication in achieving consensus with reduced communication costs. 
	As future work, an optimal selection of the number of broadcasting nodes per round could be investigated. 
	The possibility of adaptively changing $\epsilon$ could also be considered.

\end{document}